\newcommand{\half}{\ensuremath{\frac{1}{2}}}

\renewcommand{\vec}[1]{\mbox{\boldmath{$#1$}}}

\newcommand{\mstag}[1][]{\ifthenelse{\equal{#1}{}}
                            {\ensuremath{\vec{M}^\dagger}}
                            {\ensuremath{\vec{M}_{#1}^\dagger}}}
\newcommand{\mferro}[1][]{\ifthenelse{\equal{#1}{}}
                            {\ensuremath{\vec{M}^F}}
                            {\ensuremath{\vec{M}_{#1}^F}}}

\newcommand{\ham}[1][]{\ensuremath{\mathcal{\hat{H}}}}

\newcommand{\nuc}[2]{\ensuremath{\mathrm{{}^{#2} #1}}} 

\newcommand{\Curie}{\ensuremath{T_{\text C}}}
\newcommand{\et}{{\it et al}}

\documentclass[twocolumn,amsmath,amssymb,showpacs,prb]{revtex4}

\usepackage{graphicx}
\usepackage{dcolumn}
\usepackage{bm}

\begin{document}

\preprint{APS}

\title{Magnetic properties of pure and Gd doped EuO probed by NMR}

\author{Arnaud Comment}
\email{arnaud.comment@epfl.ch}
\author{Jean-Philippe Ansermet}
\affiliation{ Department of Physics and Materials Research
Laboratory, University of Illinois at Urbana-Champaign, 1110 West
Green Street, Urbana, Illinois 61801-3080 \\
and Institut de Physique des Nanostructures, Ecole Polytechnique
F\'{e}d\'{e}rale de Lausanne, CH-1015 Lausanne-EPFL, Switzerland }

\author{Charles P. Slichter}
\affiliation{ Department of Physics and Materials Research
Laboratory, University of Illinois at Urbana-Champaign, 1110 West
Green Street, Urbana, Illinois 61801-3080}

\author{Heesuk Rho}
\altaffiliation[Present address: ]{Department of Physics, Chonbuk
National University, Jeonju 561-756, Korea.}
\author{Clark S. Snow}
\author{S. Lance Cooper}
\affiliation{ Materials Research Laboratory, University of
Illinois at Urbana-Champaign, 104 South Goodwin Ave, Urbana,
Illinois 61801-3080}

\date{\today}

\begin{abstract}
An Eu NMR study in the ferromagnetic phase of pure and Gd doped
EuO was performed. A complete description of the NMR lineshape of
pure EuO allowed for the influence of doping EuO with Gd
impurities to be highlighted. The presence of a temperature
dependent static magnetic inhomogeneity in Gd doped EuO was
demonstrated by studying the temperature dependence of the
lineshapes. The results suggest that the inhomogeneity in 0.6\% Gd
doped EuO is linked to colossal magnetoresistance. The measurement
of the spin-lattice relaxation times as a function of temperature
led to the determination of the value of the exchange integral $J$
as a function of Gd doping. It was found that $J$ is temperature
independent and spatially homogeneous for all the samples and that
its value increases abruptly with increasing Gd doping.
\end{abstract}

\maketitle

\section{\label{sec:level1}Introduction}

Numerous studies on EuO have been published since the discovery of
the ferromagnetic phase of this compound by Matthias \et. in 1970,
\cite{Matthias01} and still very recently several experimental
research projects have been conducted on pure and Gd doped EuO.
\cite{Snow01,Steeneken01,Rho01,Lettieri01} There are several
reasons for this interest: first, EuO is one of the few natural
ferromagnetic semiconductors. Second, there is currently a great
deal of attention on ferromagnetic semiconductors like (Ga,Mn)As
and (In,Mn)As. \cite{Ohno01} Third, a new field called spintronics
has been developed around the possibilities of using the spin
degree of freedom of the electron in solid-state electronics.
\cite{Ball01} Fourth, EuO is an ideal system for testing new
theories in magnetism, in particular the recent developments made
on the Kondo-lattice model. The localized magnetic moments of the
half-filled $4f$-shell of the Eu atoms and the existence of a
conduction band makes EuO an appropriate system to test the
Kondo-lattice model. \cite{Schiller01,Santos01,Sinjukow01} In
addition, the low magnetic anisotropy of the material along with
the localized $J=S=7/2$ spins of the Eu$^{2+}$ ions makes europium
monoxide a nearly ideal Heisenberg ferromagnet.

Another peculiarity of EuO compounds such as O-rich EuO and Gd
doped EuO is their colossal magnetoresistance (CMR).
\cite{Shapira01,Oliver02,Petrich01,Samokhvalov02,Godart01} CMR has
also been observed in manganites. Nuclear magnetic resonance (NMR)
has been proven to be a good technique to study local magnetic
microstructures such as stripes and magnetic polarons.
\cite{Papavassiliou01,Papavassiliou02} However, to our knowledge,
such a study in europium chalcogenides has never been performed
and NMR data in EuO have been published only for temperatures far
below the transition temperature, i.e. far below the temperatures
at which CMR effect is observed in electron doped EuO. The main
reason that led us to study europium chalcogenides rather than
manganites is that the former are simple diatomic cubic crystals
whereas the structure of the latter is substantially more complex.
Consequently, studies of the (Eu,Gd)O system should allow us to
extract detailed information from NMR measurements without the
complications of numerous exotic phenomena of manganites such as
phase separation, charge and orbital ordering, and Jahn-Teller
distortions.

In this paper we present NMR results on single crystals of pure
and Gd doped EuO. In Gd doped EuO, Gd atoms play the role of
electron donors and are expected to affect the localized magnetism
of EuO negligibly. One goal of this study was to examine the
effect of Gd doping on the magnetic properties of EuO. In
Section\,\ref{sec:sample charac}, we describe the characteristics
of the samples we studied and give a detailed analysis of the NMR
lineshape of Eu in EuO. In Section\,\ref{sec:relax times}, we
describe the temperature dependence of the relaxation times of the
Eu nuclear spins. Finally, in Section\,\ref{sec:lineshape}, we
discuss the properties of the lineshape of Eu in Gd doped EuO and
discuss some models to link our observations to CMR behavior.

\section{\label{sec:sample charac}Sample characterization}

We studied four types of samples: pure EuO, 0.6\%, 2\%, and 4.3\%
Gd doped EuO. According to Samokhvalov \et., these samples can be
categorized as follows: \cite{Samokhvalov02} samples of pure and
stoichiometric EuO are insulators; samples with a Gd concentration
smaller than about 1.5\% undergo a metal-insulator transition
(MIT) when the temperature is increased above about 30\,K, the low
temperature regime being metallic; samples containing a Gd
concentration larger than about 1.5\% have a metallic behavior at
all temperatures. Godart \et. also observe similar behavior in Gd
doped EuO, \cite{Godart01} but not Schoenes and Wachter.
\cite{Schoenes01}

\subsection{Spectrometer}

The spin-spin relaxation time $T_2$ of \nuc{Eu}{151} and
\nuc{Eu}{153} nuclei in pure and Gd doped EuO can be very short,
especially when the temperature is increased towards the
transition temperature. To measure signals with short $T_2$, we
built a spectrometer with the following specifics: minimum pulse
length of 20\,ns, minimum delay between two pulses of 300\,ns,
recovery time of the receiver of 300\,ns, and maximum sampling
rate of the receiver of 1GSample/s. The high sampling rate is
needed because of the short echoes. \cite{Comment01}

In order to detect the signal using a short delay between the two
excitation pulses of a spin-echo sequence, it was necessary to use
a low Q tank circuit. Typically, we used a Q of about 10, which
corresponds to a recovery time of about 300\,ns. Note that the use
of such a small value of Q is possible because of the presence of
an amplification factor, a property inherent to magnetic
materials.

\subsection{Remarks on amplification factor}

In ferromagnetic materials, the excitation RF field $\vec{H}_1$
acting on the nuclear spins is amplified by a factor $\eta$ via
the magnetic susceptibility of the unpaired electron spins.
\cite{PetrovAndTurov} In EuO the magnetic anisotropy is low and
consequently the amplification factor for nuclei in domain walls
is not substantially stronger than for nuclei in domains. Indeed,
the amplification in domains is given by
$\eta_{Domain}=|\vec{H}_{hf}| / |\vec{H}_{an}+\vec{H}_{int}|$,
where $\vec{H}_{hf}$ is the hyperfine field, $\vec{H}_{an}$ is the
anisotropy field and $\vec{H}_{int}$ is the internal field defined
as the sum of the external field, the demagnetization field and
the Lorentz field \citep{Gossard02}. The amplification in a domain
wall of a spherical particle is given by $\eta_{DW}=\pi D
|\vec{H}_{hf}|/(N \delta |\vec{M}(T)|)$, where $N$ is the
demagnetization factor of the domain, $D$ is the domain size,
$\delta$ is the width of the domain wall and $\vec{M}(T)$ is the
magnetization in domains at temperature T,\citep{Portis01} and
therefore:

\begin{equation}
\label{Ds to DWs amp ratio}
\frac{\eta_{Domain}}{\eta_{DW}}=\frac{N}{\pi}\frac{\delta}{D}\frac{
|\vec{M}(T)|}{|\vec{H}_{an}|}.
\end{equation}

From the study of the dynamic susceptibility on an EuO sphere,
Flosdorff {\it et al.} have determined that $\delta/D\cong
0.08|\vec{M}(T=0)|/|\vec{M}(T)|$.\citep{Flosdorff01}  Then, by
taking $N=4\pi/3$, assuming a spherical domain,
$|\vec{M}(T=4.2$\,K$)|=1623$\,emu/cm$^{3}$,\citep{Samokhvalov03}
and $|\vec{H}_{an}(T=4.2$\,K$)|=247.5\pm2.5$\,Oe,\citep{Hughes01}
we obtain $\eta_{Domain}/\eta_{DW}\cong0.7$. Therefore, at 4.2\,K
the amplification in domains is only slightly weaker than the
amplification in domain walls. Moreover, since
$|\vec{H}_{an}(T)|\propto(|\vec{M}(T)|/|\vec{M}(T=0)|)^5$,\citep{Kasuya02}
the ratio (\ref{Ds to DWs amp ratio}) increases rather rapidly
with increasing temperature. It appears then that we cannot a
priori decide if the NMR signal will be dominated by nuclei in
domains or nuclei in domain walls. However, several considerations
indicate that what we observe does not depend on whether the
nuclei are located in walls or in domain walls. First, the
lineshape we observed is similar to the one observed in previous
experiments done in field high enough to suppress domain
walls.\cite{Fekete02,Lutgemeier01,Arons01} Second, in a field of
4\,T, when all domain walls are suppressed, we observed the same
temperature dependence of the relaxation times as in zero-field.
Therefore, in zero-field we did not observe any additional
relaxation mechanism related to the presence of domain walls such
as a relaxation mechanism due to domain walls motion.

\subsection{\label{sec:MvsT}Magnetization vs. temperature}

\begin{figure}
\includegraphics [width=3.41in] {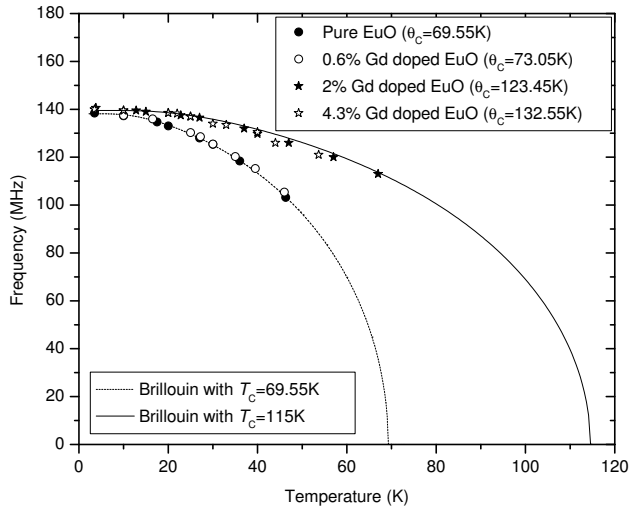}
\caption{\label{fig:fvsT} Zero-field NMR frequency of the center
of the NMR line of pure and Gd doped EuO vs. temperature.
Brillouin functions are plotted along with the data.}
\end{figure}

In ferromagnetic materials, the hyperfine field is proportional to
the magnetization. Therefore, if we assume that the nuclear
dipolar field is negligible compared to the hyperfine field, a
measurement of the temperature dependence of the zero-field
\nuc{Eu}{153} NMR frequency gives the temperature dependence of
the magnetization. The frequency measurements are presented in
Fig.\,\ref{fig:fvsT}. We determined the Curie temperature by
measuring the susceptibility $\chi$ in the paramagnetic regime,
plotting $1/\chi$ as a function of temperature and determining the
temperature at which the extrapolated $1/\chi$ line reaches zero.
This temperature corresponds to the so-called paramagnetic Curie
temperature $\theta_{\text C}$. It is important to note that,
according to the results of Samokhvalov,\cite{Samokhvalov01} in Gd
doped EuO samples the value of $\theta_{\text C}$ is larger than
the value of the Curie temperature $T_{\text C}$, the temperature
above which the spontaneous magnetization vanishes. In particular,
the value of $T_{\text C}$ of both pure and 0.6\% Gd doped EuO is
about 69.55\,K.

Although the magnetization in these systems is not well described
by mean-field theory since spin waves play a crucial role at low
temperatures, we plotted Brillouin curves along with the data as
rough approximations of the entire magnetization curves intended
to serve as guides for the eyes. The data shown in
Fig.\,\ref{fig:fvsT} are comparable to the data published by
Mauger \et.\cite{Mauger03} This is one of the indications that our
samples are consistent with those of other authors, though coming
from different sources at different times. Other such signs of
consistency among samples will come from NMR data, as shown below.

Kapusta \et. measured the temperature dependence of the frequency
in various manganites and observed that the hyperfine field did
not vanish at $T=\Curie$. \cite{Kapusta01} The authors concluded
that there was a residual magnetization above $T_{\text C}$ that
could be due to the presence of ferromagnetic regions such as
magnetic polarons. We could not observe the resonance near
$T_{\text C}$, but our data differ qualitatively since these
authors observe nearly no drop in magnetization up to $T_{\text
C}$ whereas we observed that the resonance frequencies already
decrease substantially if T is increased up to $\half T_{\text
C}$.

\subsection{\label{sec:lineshape4K}Lineshape in pure EuO at 4.2\,K}

The NMR lineshape of $\nuc{Eu}{153}$ in single crystals of pure
EuO at 4.2\,K has already been studied several times. In 1966,
Boyd found a single sharp resonance using continuous wave
NMR.\cite{Boyd01}  Later, Raj {\it et al.} observed that the
zero-field lineshape is composed of a sharp central line and two
wings but they did not discuss the broadening
mechanisms.\cite{Raj01} Then, Fekete {\it et al.} observed five
quadrupolar lines in a saturation field of 2\,T.\cite{Fekete02}
Finally, Arons {\it et al.} observed a zero-field lineshape that
they described as a sharp line on top of a broad
line.\cite{Bohn02,Lutgemeier01,Arons01} They claimed that the
sharp line corresponded to the signal from nuclei located in
domain walls and that the broad line corresponded to the signal
from nuclei in domains.

\begin{figure}
\includegraphics [width=3.41in] {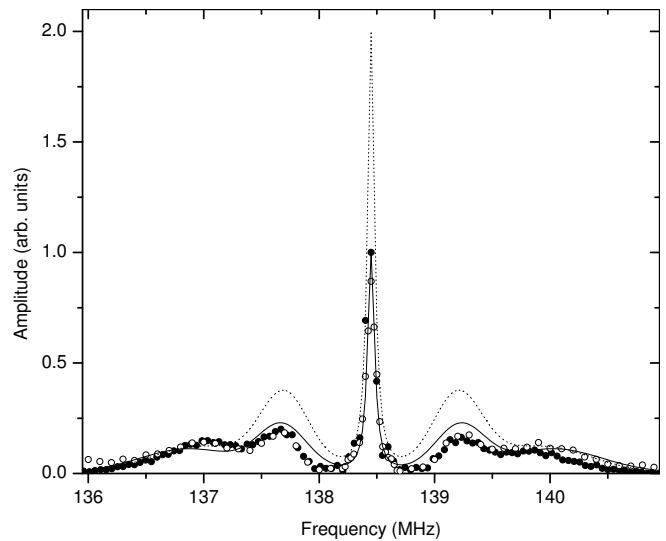}
\caption{\label{fig:Lineshape153EuEuOat4K} Zero-field lineshape of
$\nuc{Eu}{153}$ in EuO at 4.2\,K: the black dots correspond to the
FT of the echo and the open circles to the point-by-point
measurement. The dotted line is the fit of the data using
(\ref{Lineshape}) and the plain line is the same fit corrected for
$T_2$ effects.}
\end{figure}

We present in Fig.\,\ref{fig:Lineshape153EuEuOat4K} the zero-field
lineshape of $\nuc{Eu}{153}$ in pure EuO that we measured at
4.2\,K. We determined the lineshape by two different techniques:
on one hand, we measured the echo integral at a certain number of
discrete values of frequency (point-by-point measurement).
Alternatively, we performed a Fourier transform (FT) of the echo
measured at the frequency of the central line. As shown in
Fig.\,\ref{fig:Lineshape153EuEuOat4K}, both methods give an almost
identical lineshape.

\begin{figure}
\includegraphics [width=3.41in] {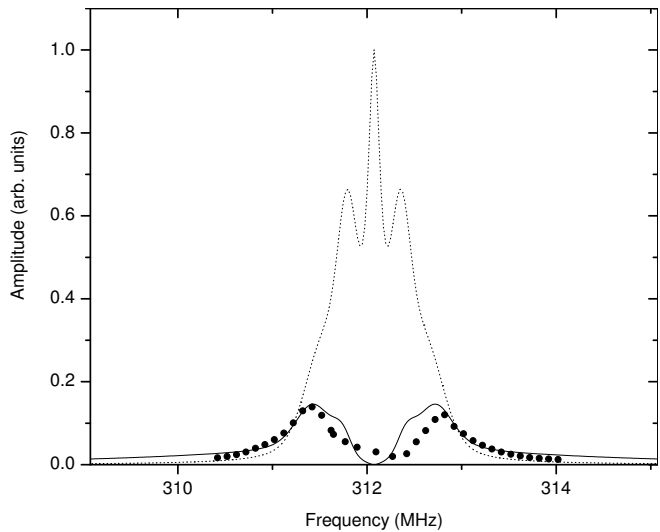}
\caption{\label{fig:Lineshape151EuEuOat4K} Zero-field lineshape of
$\nuc{Eu}{151}$ in EuO at 4.2\,K: the black dots correspond to the
point-by-point measurement. The dotted line is the computed
lineshape and the plain line is the same computed lineshape
corrected for $T_2$ effects.}
\end{figure}

Since contradictory results and analysis of the lineshape of
$\nuc{Eu}{153}$ in pure EuO at 4.2\,K were published, we present
here our own analysis. Because the Eu sites in EuO have a nominal
cubic symmetry, we expected to observe a single narrow line.
However, as shown in Fig.\,\ref{fig:Lineshape153EuEuOat4K}, we
observed an intense central peak and two wings with a structure
suggesting that each wing is composed of two broadened lines. In
order to distinguish between the broadening due to electric field
gradients (EFGs) and the broadening due to the magnetic
environment of the nuclei, we took advantage of the fact that
there are two isotopes of europium with similar natural abundance.
The gyromagnetic ratio and the electric quadrupole moment of the
two isotopes are shown in Table\,\ref{tab:europium isotopes}. The
zero-field lineshape of $^{151}$Eu in pure EuO at 4.2\,K is shown
in Fig.\,\ref{fig:Lineshape151EuEuOat4K}. By comparing the
$^{153}$Eu and the $^{151}$Eu line at 4.2\,K we observed that the
lines cannot be described solely by magnetic broadening or by
quadrupole electrical broadening. We therefore came to the
conclusion that the broadening was due to a combination of both
effects. We will discuss the origin of such a broadening in
Sect.\,\ref{sec:temp dependance lineshape}.

\begin{table}
\caption{\label{tab:europium isotopes} Spin, natural abundance,
gyromagnetic ratio, and electric quadrupole moment of the two
europium isotopes.\cite{CRC01}}
\begin{ruledtabular}
\begin{tabular}{ccccc}
{} & Spin & Nat. abund. (\%) & $\gamma_n/2\pi$ (MHz/T) & $Q$ (barn)\\
\hline
$\nuc{Eu}{153}$ & 5/2 & 52.19 & 4.6745 & 2.41\\
$\nuc{Eu}{151}$ & 5/2 & 47.81 & 10.5856 & 0.903\\
\end{tabular}
\end{ruledtabular}
\end{table}

In the presence of EFGs, since $\nuc{Eu}{153}$ has a spin $I=5/2$,
we expected to observe five lines all separated by the same
frequency interval $\Delta\nu_{Q}$ and with intensity ratios
5:8:9:8:5.\cite{Slichter01} We tried to reproduce the observed
lineshape theoretically by assuming a distribution in EFGs and a
magnetic broadening. We described the inhomogeneities in the EFGs
with a Gaussian distribution leading to a distribution in
frequency $G(\nu)$ of second moment $\delta_{\Delta\nu_{Q}}$ that
broadens the five lines. As will be discussed in
Section\,\ref{sec:T2}, a Lorentzian distribution describes
correctly the magnetic broadening. The broadening resulting from
these two mechanisms is a convolution of the two distribution
functions. The convolution of a Gaussian and a Lorentzian gives a
Voigt function, i.e., a function of the form

\begin{equation}
\label{Voigt}
V(\nu)=\frac{\frac{1}{2}\Gamma}{\pi^{3/2}}\int_{-\infty}^{\infty}\frac{e^{-t^2}}{(\nu-\nu_{0}-\sqrt{2}\sigma
t)^2+(\frac{1}{2}\Gamma)^2}\,dt,
\end{equation}
where $\Gamma$ is the full width at half maximum (FWHM) of the
Lorentzian and $\sigma$ is the second moment of the
Gaussian.\cite{Thompson01} The resulting lineshape is then given
by

\begin{figure}
\includegraphics [width=3.41in] {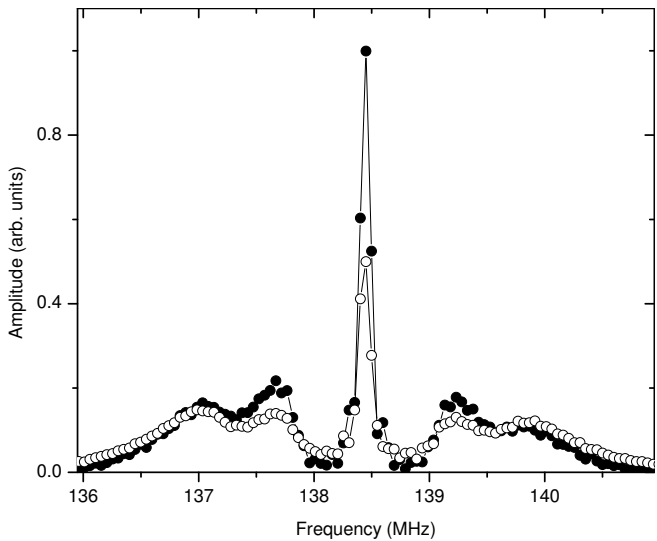}
\caption{\label{fig:2DNMR} Zero-field lineshape of $\nuc{Eu}{153}$
in EuO at 4.2\,K obtained by FT. The black dots correspond to a
delay of 10.5\,$\mu$s and the open circles to a delay of
15.5\,$\mu$s. This shows the presence of frequency dependent
$T_2$'s (the $T_2$'s are shorter near the line center). }
\end{figure}

\begin{widetext}
\begin{equation}
\label{Lineshape} f(\nu)=\sum_{m=-2}^{m=2}(1-\frac{m^{2}}{9})
\frac{\frac{1}{2}\Gamma}{\pi^{3/2}}\int_{-\infty}^{\infty}\frac{e^{-t^2}}{(\nu-(\nu_{0}-m\Delta\nu_{Q})-\sqrt{2}|m|\delta_{\Delta\nu_{Q}}
t)^2+(\frac{1}{2}\Gamma)^2}\,dt,
\end{equation}
\end{widetext}
where $\Gamma$ represents the FWHM of the magnetic distribution
and the factor $(1-m^2/9)$ accounts for the difference in
intensity of the five quadrupolar lines. By fitting the measured
$\nuc{Eu}{153}$ lineshape with this function, it was possible to
determine the parameters $\Gamma$, $\Delta\nu_{Q}$ and
$\delta_{\Delta\nu_{Q}}$. The fit is represented by a dotted line
in Fig.\,\ref{fig:Lineshape153EuEuOat4K} and the deduced values of
the fitting parameters are shown in Table \ref{tab:fit
parameters}. However, there was another characteristic that needed
to be taken into account: Raj {\it et al.} showed that the
$\nuc{Eu}{153}$ spin-spin relaxation time in pure EuO is short and
frequency dependent and, as a consequence, some of the nuclei
might not be observed if the delay between the excitation pulses
is too long.\cite{Raj01} They observed that the intensity of the
central part of the line is strongly reduced. We demonstrated this
phenomena in Fig.\,\ref{fig:2DNMR} where we plotted the FT of the
echo measured with two different delays.\footnote{Since the
central peak is sharp, the echo is long and it was not possible to
use delays shorter than about 10\,$\mu$s to record the entire
echo, even with the short recovery time of 0.3\,$\mu$s of the
receiver.} We took this effect into account in the computation of
the lineshape of $\nuc{Eu}{153}$ by multiplying (\ref{Lineshape})
by an approximated shape of the frequency distribution of $T_2$ at
4.2\,K deduced from our measurements. We will discuss the physical
reason for this $T_{2}$ distribution in Sect.\,\ref{sec:T2}. The
computed lineshape is shown in plain line in
Fig.\,\ref{fig:Lineshape153EuEuOat4K} along with the measured
lineshape.

\begin{table}
\caption{\label{tab:fit parameters} Values of the lineshape
fitting parameters for $\nuc{Eu}{153}$ along with the deduced
parameters for $\nuc{Eu}{151}$.}
\begin{ruledtabular}
\begin{tabular}{cccc}
{} & $\Gamma$ [MHz] & $\Delta\nu_{Q}$ [MHz] & $\delta_{\Delta\nu_{Q}}$ [MHz]\\
\hline
$\nuc{Eu}{153}$ & 0.072 & 0.75 & 0.2\\
$\nuc{Eu}{151}$ & 0.163 & 0.281 & 0.075\\
\end{tabular}
\end{ruledtabular}
\end{table}

To verify if our description of the $\nuc{Eu}{153}$ lineshape was
correct, we computed the $\nuc{Eu}{151}$ lineshape from the
fitting parameters that we determined for $\nuc{Eu}{153}$ and
compared the resulting curve with the measured $\nuc{Eu}{151}$
lineshape. We first calculated the parameters for $\nuc{Eu}{151}$
using the relations

\begin{eqnarray}
\label{magnetic width ratio}
\Gamma_{151} &=& \frac{^{151}\gamma_n}{^{153}\gamma_n}\;\Gamma_{153}, \\
\Delta\nu_{Q,151} &=& \frac{^{151}Q}{^{153}Q}\;\Delta\nu_{Q,153}, \\
\delta_{\Delta\nu_{Q},151} &=&
\frac{^{151}Q}{^{153}Q}\;\delta_{\Delta\nu_{Q},153},
\end{eqnarray}
where the indices 151 and 153 refer to $\nuc{Eu}{151}$ and
$\nuc{Eu}{153}$ respectively. The computed lineshape is shown as a
dotted line in Fig.\,\ref{fig:Lineshape151EuEuOat4K}. Then, we
deduced the $T_2$ distribution of $\nuc{Eu}{151}$ from the $T_2$
distribution of $\nuc{Eu}{153}$ using the following process:
first, since we observed that $T_2$ at 4.2\,K was inversely
proportional to $\gamma_n^2$ (c.f. Section\,\ref{sec:T2}), we
divided the amplitude of the distribution by
$(^{151}\gamma_n/^{153}\gamma_n)^2$. Second, since $T_2$ depends
on $\gamma_n$ (and is therefore of magnetic origin), we expected
the $T_2$ distribution to scale with the magnetic distribution. We
thus calculated the $\nuc{Eu}{151}$ $T_2$ distribution by
multiplying the width of the $\nuc{Eu}{153}$ $T_2$ distribution by
$^{151}\gamma_n/^{153}\gamma_n$. The deduced distribution is in
good agreement with our data. In order to take into account the
fact that the $T_2$ of the central peak was too short to be
detected, we introduced a cutoff in the computed distribution
curve. Finally, the $T_2$ corrected $\nuc{Eu}{151}$ lineshape was
obtained by multiplying the computed lineshape by the resulting
$T_2$ distribution. The result is plotted in solid line in
Fig.\,\ref{fig:Lineshape151EuEuOat4K} along with the measured
data. The calculated line was in reasonable agreement with the
data, thus confirming our description of the lineshape in terms of
a magnetic and a quadrupolar broadening.\footnote{From the above
analysis, we deduced a ratio $^{151}\gamma_n/^{153}\gamma_n$
slightly different than the one given in tables. Indeed, by
comparing the frequency of the central transition of both
isotopes, we obtained a ratio of about 2.254 and the value given
in tables is about 2.2645.\cite{CRC01}}

The origin of the quadrupolar splitting is most likely intrinsic
to EuO and not due to oxygen vacancies as suggested by Arons {\it
et al.}\cite{Arons03} Oxygen vacancies will lead to a clear
breaking of the cubic symmetry and therefore, since the
quadrupolar moment of Eu is large, the quadrupolar splitting is
expected to be large and the Eu sites located next to the
vacancies will not contribute to the observed line. Note that the
electric quadrupole splitting we observed in zero-field is similar
to the one observed by Fekete \et. in an external field
$\vec{H}_0$ along the easy axis [111] saturating the magnetization
(2\,T$<|\vec{H}_0|<$3\,T).\cite{Fekete02}

\section{\label{sec:relax times}Effects of magnons and doping}

The measurement of NMR relaxation times in europium chalcogenides
gives some valuable information about the dynamics of their
electron spins through the hyperfine coupling. Boyd was the first
to report NMR data on EuS, \cite{Boyd02} and together with Charap,
used spin-wave theory to deduce values of $J_1$ and $J_2$,
respectively the nearest neighbor and next nearest neighbor
exchange constant, from the temperature behavior of the NMR
frequency of both $\nuc{Eu}{151}$ and
$\nuc{Eu}{153}$.\cite{Charap01} The first values of NMR relaxation
times in EuO were published in 1965 by Uriano and Streever.
\cite{Uriano01} Following this work, several groups studied NMR
relaxation times and compared their data to spin-wave theory and
to the Suhl-Nakamura theory in order to explain their results and
to obtain information on the magnetic properties of the system.
All the results published concern measurements performed at
$T\leq20.3$\,K. We extended the study of relaxation times to much
higher temperatures and to Gd doped EuO. We determined the
influence of doping on the dynamic properties of the electronic
magnetization by comparing the relaxation times in EuO to the
relaxation times in Gd doped EuO.

\subsection{\label{sec:T1}Spin-lattice relaxation times}

We used a saturation recovery sequence to perform $T_1$
measurements.  The sequence destroys the nuclear magnetization by
applying a series of pulses before measuring the signal. It
consists of a series of pulses of $0.2\mu s$ separated by a time
of the order of $T_2$ forcing the nuclear spins to loose their
coherence. A standard spin-echo sequence is applied after a while
to detect the amplitude of the magnetization that has recovered
equilibrium. Note that, unlike for $T_2$ measurements, no
substantial frequency dependence of $T_1$ was observed, neither in
pure EuO nor in Gd doped EuO.

\begin{figure}
\includegraphics [width=3.41in] {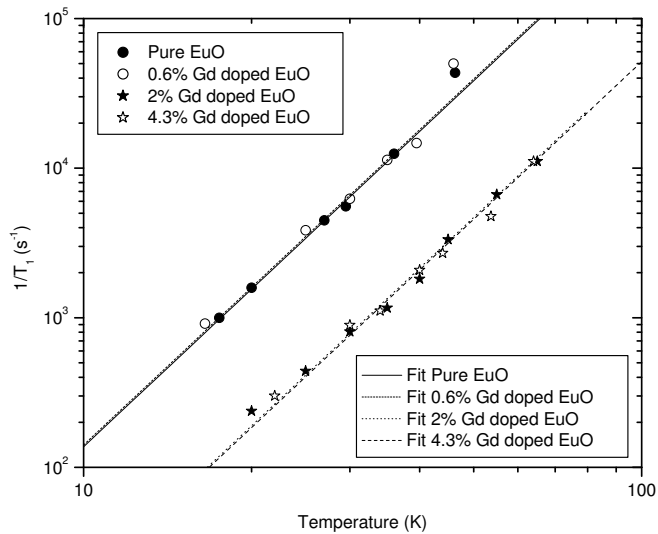}
\caption{\label{fig:T1vsT} Zero-field spin-lattice relaxation
rates of $\nuc{Eu}{153}$ in pure and Gd doped EuO as a function of
temperature. The expression used to fit the curves is defined in
(\ref{Oguchi}).}
\end{figure}

Barak, Gabai and Kaplan determined that the spin-lattice
relaxation in powdered EuO in an external magnetic field at
temperatures below 14\,K are dominated by two-magnon
processes.\cite{Barak01} Also, between 14\,K and 20\,K, they
observed that a three-magnon process surpasses the two-magnon
process and that it is enhanced by a second-order three-magnon
process. The results we obtained at zero-field confirmed that, for
temperatures above about 14\,K, the three-magnon process is the
dominant spin-lattice relaxation process in EuO. The relaxation
rate (Fig.\,\ref{fig:T1vsT}) was found proportional to $T^{7/2}$.
This is consistent with the temperature dependence of the
theoretical expression first derived by Oguchi and Keffer for a
three-magnon process.\cite{Oguchi01} We rederived the expression
and adapted it to the case of zero-field measurements to obtain
(Appendix \ref{sec:appendixA}):

\begin{equation}
\label{Oguchi}
\frac{1}{T_{1}}=\frac{11.29}{16(2\pi)^5}\frac{A^2}{2JS\cdot \hbar
S}\left(\frac{k_{B}T}{2JS}\right)^{7/2},
\end{equation}
where $A$ is the hyperfine constant, $J=J_1+J_2$ the exchange
integral, $S$ the electron spin, and $k_B$ the Boltzmann constant.
As for the results of Barak \et., we had to multiply the
expression (\ref{Oguchi}) by the exchange scattering enhancement
factor that we determined to be $\xi\cong8$, in agreement with the
calculation of Beeman and Pincus.\cite{Beeman01} Taking $S=7/2$
and determining $A$ from the zero-field resonance frequency
extrapolated to $T=0$\,K,
$A=-\hbar\omega/S\cong-2.6\cdot10^{-26}$\,J, we deduced from the
fit a value of the exchange integral of $J/k_{B}=0.755\pm0.01$\,K.

NMR thus provides an accurate determination of $J$. Even if the
enhancement factor is not precisely known, the fact that $1/T_1$,
and therefore $\xi/T_1$, is proportional to $J^{-9/2}$ (c.f.
(\ref{Oguchi})) means that a large change in $\xi$ leads to a
small change in $J$. The influence of a variation in the
enhancement factor on the determination of $J$ is therefore
limited, and we evaluated the error to be less than 1\%. The other
causes of uncertainty are the error on the $T_1$ measurement and
the error on the temperature measurement. An evaluation of the
total error leads to $\Delta J/k_B=0.01$\,K.

Previous measurements on powdered EuO by neutron scattering
experiments and specific heat measurements led respectively to
$(J_1+J_2)/k_B=0.725\pm 0.006$\,K,\cite{Passell01} and
$(J_1+J_2)/k_B=0.714\pm 0.007$\,K.\cite{Dietrich01} More recently,
Mook measured single crystals of EuO by neutron scattering methods
and he obtained the following values: $J_1/k_B=0.625\pm 0.007$\,K
and $J_2/k_B=0.125\pm 0.01$\,K.\cite{Mook01} These values lead to
a value of the exchange integral $J=0.750\pm0.017$\,K which is in
very good agreement with our measurement. Note that in 1966 E. L.
Boyd deduced $J_1$ and $J_2$ by measuring the temperature
dependence of the NMR frequency of $\nuc{Eu}{153}$.\cite{Boyd01}
He found a value of $J_1/k_B=0.750\pm0.0025$\,K that is very close
to the value of $J$ that we determined. However, he deduced a
negative value of $J_2$.

In Gd doped EuO (Fig.\,\ref{fig:T1vsT}), we observed that the
temperature dependence of the spin-lattice relaxation times is the
same $T^{7/2}$. Hence, as for pure EuO, the three-magnon process
is the dominant relaxation process. Using expression
(\ref{Oguchi}) and the enhancement factor $\xi=8$, we deduced the
values of $J$ shown in Table \ref{tab:Jvalues} for the doped
samples. Clearly, the samples can be separated in two categories
according to their value of $J$ (Fig.\,\ref{fig:T1vsT}).

To the best of our knowledge, no direct measurement of the
exchange integral in Gd doped EuO has ever been published and our
results provide therefore the first measurement of $J$ in Gd doped
EuO.

\begin{table}
\caption{\label{tab:Jvalues}Values of J as a function of Gd
doping. }
\begin{ruledtabular}
\begin{tabular}{cc}
Doping level x (\%) & Exchange integral $J/k_B$ (K) \\
\hline
0 & $0.755 \pm 0.01$ \\
0.6 & $0.750\pm0.01$ \\
2 & $1.205\pm0.01$ \\
4.3 & $1.210\pm0.01$ \\
\end{tabular}
\end{ruledtabular}
\end{table}

\subsection{\label{sec:T2}Spin-spin relaxation times}

We performed spin-spin relaxation measurements in pure EuO and in
Gd doped EuO using a standard spin-echo sequence composed of two
consecutive pulses of duration $t_1$ and $t_2$ separated by a
varying delay. Typical values were $t_1=0.1$\,$\mu$s and
$t_2=0.2$\,$\mu$s. Since $T_2$ is strongly frequency dependent, it
is necessary to specify at what frequency it was measured. In the
following, $T_2$ always corresponds to the spin-spin relaxation
time measured at the frequency corresponding to the center of the
central line, or the center of gravity of the line in case the
central line is not well defined. Note that all the relaxation
times that will be presented in the following correspond to
measurements done with short delays. Therefore, the discussion and
the analysis will focus on the fast decay times unless specified
otherwise.

Except for the value of $T_2$ at 20.3\,K determined by Uriano and
Streever \cite{Uriano01} and the decay curve at 13.8\,K published
by Barak {\it et al.}, \cite{Barak02} there are no spin-spin
relaxation time data for EuO above 4.2\,K in the literature.
However, several studies of $T_2$ at 4.2\,K and lower temperatures
were performed following the work of Barak {\it et al.}:
\cite{Barak03,Barak02} Raj {\it et al.},\cite{Raj01} Fekete {\it
et al.}, \cite{Fekete01} the L\"{u}tgemeier
group,\cite{Lutgemeier01,Arons01,Bohn01,Arons02,Arons03} and more
recently Pieper {\it et al.}\cite{Pieper01} The main reason for
this interest is that the decay curve is not a single exponential.
As L\"{u}tgemeier {\it et al.}\cite{Lutgemeier01} first noted,
this is because there are two relaxation mechanisms: a fast one
due to the Suhl-Nakamura indirect interaction, and a slow one due
to direct dipolar coupling. However, the original Suhl-Nakamura
theory fails to explain the observed spin-spin relaxation time and
its frequency dependence that was first reported by Raj {\it et
al.}.\cite{Raj01} Barak {\it et al.} used the theory developed by
Hone, Jaccarino, Ngwe and Pincus \cite{Hone01}(in the following we
will refer to this theory as the HJNP theory) to explain why the
value of $T_2$ they observed in a powdered EuO sample at 4.2\,K is
longer than the one predicted by the Suhl-Nakamura
theory.\cite{Barak02} The HJNP theory assumes the existence of an
inhomogeneous line broadening resulting from random microscopic
inhomogeneities and predicts a frequency dependent $T_2$. The HJNP
theory predicts a Lorentzian line shape and for this reason we
used a Lorentzian as magnetic distribution function in the
calculation of the lineshape presented in
Section\,\ref{sec:lineshape4K}.

In 1976, Fekete {\it et al.} successfully described the decay
curves of the five transitions they had observed by assuming an
inhomogeneous Suhl-Nakamura relaxation process (deduced from the
HPJN model) and a dipolar coupling.\cite{Fekete01} They took into
account all the elements necessary to describe the lineshape and
its variation with delay as shown in Fig.\,\ref{fig:2DNMR}.
However, they did not discuss the implication of their results on
the lineshape. Arons {\it et al.} published in 1975 the zero-field
lineshape vs. delay and observed, as we did, the reduction of the
amplitude of the central peak with increasing delay.
\cite{Arons01} However, Arons {\it et al.} claimed that the $T_2$
of the central line was shorter because the signal was coming from
nuclei located in domain walls. Our combined results on $T_1$ and
$T_2$ measurements as a function of temperature point to more
intrinsic mechanisms.

\begin{figure}
\includegraphics [width=3.41in] {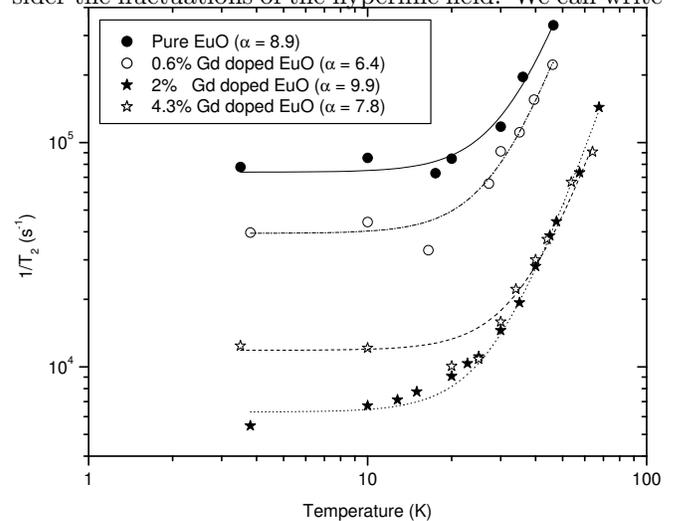}
\caption{\label{fig:T2vsT} Zero-field spin-spin relaxation rates
of $\nuc{Eu}{153}$ in pure and Gd doped EuO as a function of
temperature. The parameter $\alpha$ and the fitting curve are
defined in (\ref{T2Redfield}).}
\end{figure}

We measured the temperature dependence of the spin-spin relaxation
times (Fig.\,\ref{fig:T2vsT}). For pure EuO as well as for all the
Gd doped samples, $T_2$ seems to be temperature independent at
temperatures below about 15\,K. For higher temperatures, we
observed a rapid increase of the relaxation rates with increasing
temperature. We analyze these two regimes below.

\subsubsection{Lifetime effect}

We saw in Sect.\,\ref{sec:T1} that the spin-lattice relaxation
processes are dominated by the scattering of magnons by nuclear
spins, in particular by processes involving three magnons. One way
to describe these processes is to consider the fluctuations of the
hyperfine field. We can write the perturbation Hamiltonian as

\begin{equation}
\label{fluctuations} \ham^{pert}_{hf}(t)=-\gamma_n \hbar
[H_x(t)\hat{I}_x+H_y(t)\hat{I}_y+H_z(t)\hat{I}_z],
\end{equation}
where $H_x(t)$, $H_y(t)$ and $H_z(t)$ are the fluctuating fields
at the nuclear site due to fluctuations of $\vec{\hat{S}}$ in the
$x$, $y$ and $z$ directions respectively.\footnote{We do not need
to make any assumption on the anisotropy of the hyperfine
coupling. This expression is valid even if the hyperfine coupling
needs to be describe by a tensor.} The relaxation rates deriving
from these interactions can be calculated using the Redfield
theory.\cite{Slichter01} The calculation for the particular case
of a spin $I=\frac{5}{2}$ gives the following result for the
spin-spin relaxation time:

\begin{equation}
\label{T2Redfield} \frac{1}{T_2}=\gamma_n^2
\overline{H_z^{2}}\tau_{0}+\alpha \frac{1}{T_1},
\end{equation}
where $\overline{H_z^2}$ is the amplitude of the correlation
function between $H_z(t)$ and $H_z(t+\Delta t)$, $\tau_0$ is the
correlation time (or lifetime) of the scattering process between
the magnons and the nuclear spins, and

\begin{widetext}
\begin{equation}
\label{Alpha values} \alpha=I(I+1)-m(m+1) = \left\{
\begin{array} {lll}
    5 & \text{for the $-5/2\leftrightarrow -3/2$ and $3/2\leftrightarrow 5/2$ transitions} \\
    8 & \text{for the $-3/2\leftrightarrow -1/2$ and $1/2\leftrightarrow 3/2$ transitions} \\
    9 & \text{for the $-1/2\leftrightarrow 1/2$ transition}.
\end{array} \right.
\end{equation}
\end{widetext}
We have assumed that the fluctuations of the three components of
field $x$, $y$, and $z$ are independent and that the correlation
functions are simple exponential. \cite{Slichter01} Since the
length of the pulses as well as the amplitude of the RF field were
chosen such as to excite all the transitions, we expect to have a
mixture of these rates. As shown in Fig.\,\ref{fig:T2vsT}, the
factor $\alpha$ obtained from fitting the curves with the function
$1/T_2=\alpha/T_1+\beta$ is between 6.3 and 9.9. This is in good
agreement with the theoretical values (\ref{Alpha values}).
Therefore, according to the Redfield theory, the temperature
dependence of the spin-spin relaxation in pure and Gd doped EuO is
entirely determined by transverse fluctuating fields. Our results
showed that these fluctuations were due to fluctuations of the
electronic spins well described by spin-wave theory.

\subsubsection{Temperature independent mechanisms}

For temperatures far below the magnetic transition temperature,
the Suhl-Nakamura interaction is expected to be temperature
independent since its temperature dependence comes mostly from the
hyperfine constant.\cite{Suhl01,Nakamura01} The nuclear
dipole-dipole interaction is also independent of temperature.
Therefore, two processes compete and may be the source of the
observed temperature independent $T_2$'s. In the case of pure EuO,
the relaxation time due to the dipole-dipole interaction is
considerably slower and we observed, in agreement with previous
measurements, that the Suhl-Nukamura processes dominate the
dipole-dipole processes for short delays. As shown in
Fig.\,\ref{fig:T2vsT}, the relaxation time of $\nuc{Eu}{153}$ in
Gd doped EuO is substantially slower than in pure EuO. In 2\% Gd
doped EuO at 3.8\,K, we have measured $T_2=183$\,$\mu$s. This
value is very close to the value $T_2=194\pm2.4$\,$\mu$s deduced,
following the treatment of Bohn {\it et al.},\cite{Bohn01} from an
exponential fit of the slow spin-echo decay measured in pure EuO
and thought to be due to dipole-dipole interactions.\footnote{The
dipole-dipole relaxation time in Gd doped EuO is expected to be
slightly longer than in pure EuO due to the fact that interactions
between Eu nuclear spins and Gd nuclear spins (unlike spins) lead
to less effective relaxation processes. \cite{Slichter01}}

We observed that adding Gd in the EuO matrix leads to magnetic
broadening, which increases with doping (c.f. Section
\ref{sec:lineshape}). Therefore, in agreement with the HJNP
theory, the replacement of Eu atoms by Gd atoms in the EuO matrix
reduces the allowed mutual spin-flips between Eu nuclei leading to
the Suhl-Nakamura relaxation and consequently we expect the
relaxation rates to be smaller when the concentration of Gd is
increased. The relaxation through dipole-dipole interactions is
however still effective since in addition to the vanishing
$\hat{I}^+_i\hat{I}^-_j$ terms, the interaction Hamiltonian has
diagonal $\hat{I}_{z,i}\hat{I}_{z,j}$ terms that do not vanish
even if the frequencies of spins $i$ and $j$ are very different.
Note that the temperature independent relaxation rate is larger
for 4.3\% Gd doped EuO than for 2\% Gd doped EuO. The presence of
an increasing number of conduction electrons when the Gd
concentration is increased above about 1.5\% might introduce an
additional relaxation process through an RKKY indirect
interaction.

\section{\label{sec:lineshape} Magnetic inhomogeneities}

\subsection{\label{sec:temp dependance lineshape}Temperature dependence of the pure EuO lineshape}

\begin{figure}
\includegraphics [width=3.41in] {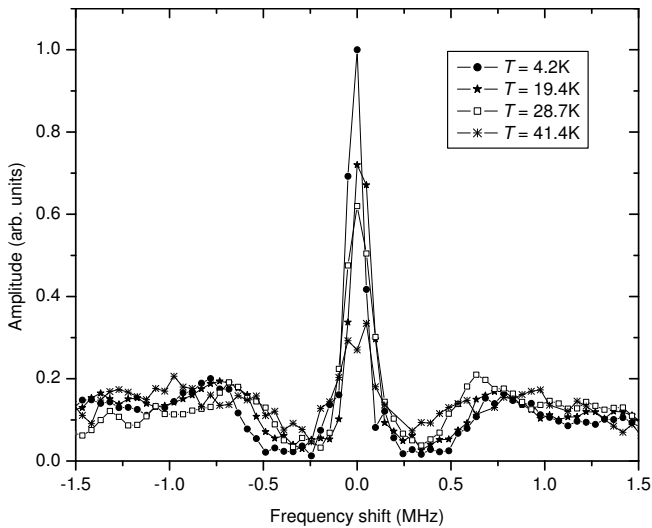}
\caption{\label{fig:LineshapeEuOvsT}Zero-field lineshape of
$\nuc{Eu}{153}$ in EuO vs. temperature. The lineshapes were
obtained by Fourier transform of the spin-echo signal.}
\end{figure}

To the best of our knowledge, no measurement of the lineshape of
pure EuO above 4.2\,K has ever been published. We report here the
first study of the temperature dependence of the $\nuc{Eu}{153}$
lineshape. In Fig.\,\ref{fig:LineshapeEuOvsT}, we present the
lineshape of pure EuO at 4.2\,K and at several temperatures
between 4.2\,K and 42\,K. These lines were obtained by Fourier
transforming the spin-echo measured at the frequency of the center
of the lines. Since, as shown in Section\,\ref{sec:MvsT}, the NMR
frequency varies considerably with the temperature, we shifted the
center of the lines presented in the figure to zero frequency in
order to compare the lines recorded at different temperatures. We
also adjusted the amplitude of the lines to superimpose the wings
of all the lines. This adjustment allowed us to compare the
amplitude and the width of the central peak of the lines.

We observed that the intensity of the central peak was greatly
reduced with increasing temperature. As noted in
Section\,\ref{sec:T2}, the spin-spin relaxation times are
frequency dependent and they are shorter for frequencies close to
the central peak. Also, we observed that $T_2$ is strongly
temperature dependent, decreasing with increasing temperature.
Therefore, we concluded that the intensity of the central peak
decreased with increasing temperature as a consequence of the
decreasing value of $T_2$, our spectrometer not being able to
detect all the nuclei of the central peak. We also observed that
the width of the central peak increases almost linearly from about
0.11\,MHz at 4.2\,K to about 0.23\,MHz at 41.4\,K. This broadening
mechanism did not seem to be linked to $T_2$ effects since we
measured $T_2\cong4$\,$\mu$s at 41.4\,K, which corresponds to a
Lorentzian width of about 0.08\,MHz.

Finally, we observed that the electrical quadrupole splitting
decreases with increasing temperature. We estimated the change to
be of the order of 25\%, from $\Delta\nu_{Q}\cong0.75$\,MHz at
4.2\,K to $\Delta\nu_{Q}\cong0.55$\,MHz at 41.4\,K. We think that
this change might be due to a variation of the magnetostriction
with temperature. This assumption is supported by the results of
F. Levy who observed, by X-ray measurements, that the spontaneous
magnetostriction of EuO decreases by about 25\% between 4.2\,K and
40\,K.\cite{Levy01} Note that if the quadrupolar splitting was due
to the presence of defects in the crystals, as was proposed by
Arons {\it et al.},\cite{Arons03} we would probably not observe
such a temperature dependence.

\subsection{Influence of Gd on EuO lineshape}

\begin{figure}
\includegraphics [width=3.41in] {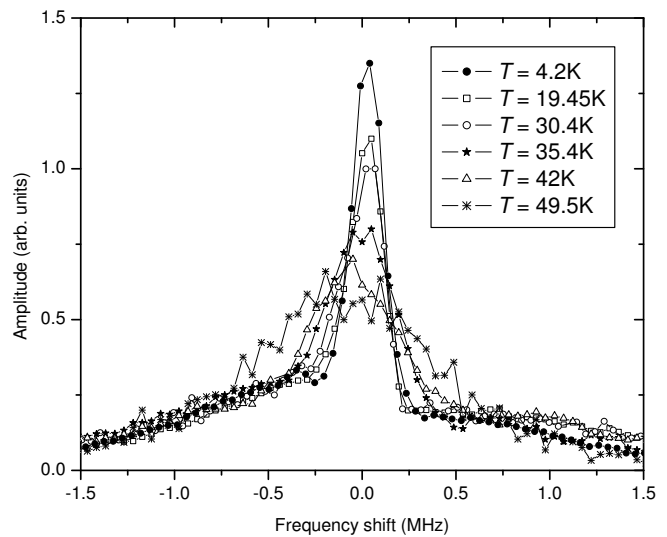}
\caption{\label{fig:LineshapeGd6EuOvsT}Zero-field lineshape of
$\nuc{Eu}{153}$ in 0.6\% Gd doped EuO vs. temperature. The
lineshapes were obtained by Fourier transform of the spin-echo
signal.}
\end{figure}

We now turn to the analysis of the lineshapes of $\nuc{Eu}{153}$
in Gd doped EuO. We observed that the frequency of the central
transition of the 0.6\% Gd doped EuO lineshape at 4.2\,K was only
slightly shifted towards higher frequency compared to pure EuO,
from 138.45\,MHz to 138.48\,MHz. However, the central line was
substantially broader as can be seen on
Fig.\,\ref{fig:LineshapeGd6EuOvsT}. As discussed in
Section\,\ref{sec:lineshape4K}, the width of the central
transition is determined by magnetic broadening only. We think
that the observed line at 4.2\,K is broader in the 0.6\% sample
than in the pure sample because of the random distribution of Gd
atoms in the EuO matrix. Since different Eu atoms have different
numbers of nearest and next nearest Gd neighbors, the local field
acting on Eu atoms, which is influenced by the presence of Gd, is
not the same at each Eu site. An other noticeable difference
between pure EuO and 0.6\% Gd doped EuO is that the quadrupolar
structure is not resolvable in 0.6\% Gd doped EuO. We think that
this is also due to the random distribution of Gd atoms in EuO
which leads
to a random distribution of EFGs. 

\begin{figure}
\includegraphics [width=3.41in] {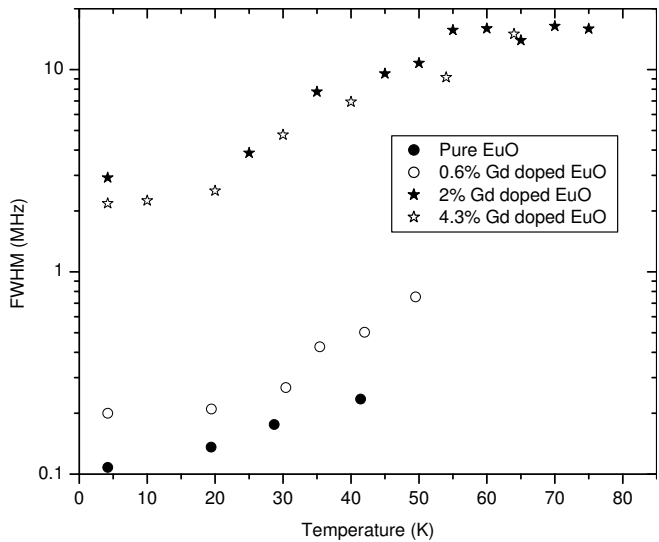}
\caption{\label{fig:FWHMvsT} FWHM of the zero-field lineshape of
Eu$^{153}$ in pure EuO and 0.6\%, 2\% and 4.3\% Gd doped EuO vs.
temperature.}
\end{figure}

We also measured the lineshape of 0.6\% Gd doped EuO as a function
of temperature from 4.2\,K to 49.5\,K
(Fig.\,\ref{fig:LineshapeGd6EuOvsT}). We observed a temperature
dependent broadening of the central line. We adjusted the
amplitude of the different lines by superimposing the wings of all
the lines.\footnote{While the intensity of the central line in
pure EuO was decreasing with increasing temperature due to short
$T_2$'s, we observed that the area under the central peak of 0.6\%
Gd doped EuO was to a good approximation temperature independent.
This is most likely because the spin-spin relaxation times in
0.6\% Gd doped EuO are slower by a factor of about 1.5.} We
plotted in Fig.\,\ref{fig:FWHMvsT} the full width at half maximum
(FWHM) of the central line as a function of temperature for all
the samples.\footnote{We defined the half maximum of the central
line of 0.6\% Gd doped EuO from the top of the wings and not from
zero intensity. Therefore, we probably underestimated the FWHM of
the lineshape of 0.6\% Gd doped EuO.} It appears that for
temperatures above about 30\,K, the FWHM of the 0.6\% Gd doped EuO
lineshape increases sharply with increasing temperature. We did
not observed this phenomena in pure EuO. We demonstrate that the
broadening could not be explained by a variation in $T_2$ and that
the broadening was therefore a static broadening. Indeed, the
measured $1/T_2$ is much less than the observed linewidth.
Moreover, we observed that $T_2$ has the same temperature
dependence in 0.6\% Gd doped EuO as in pure EuO (c.f.
Section\,\ref{sec:T2}), and the spin-spin relaxation rates in
0.6\% Gd doped EuO are longer than in pure EuO.

The addition of 0.6\% Gd in the EuO matrix gives rise to
temperature dependent phenomena associated with a static magnetic
inhomogeneity. The temperature dependent broadening mechanism
initiates at about 30\,K, the temperature at which the
conductivity starts decreasing dramatically according to the
results of Godart {\it et al.} \cite{Godart01} and Samokhvalov
{\it et al.} \cite{Samokhvalov02} This suggests that the phenomena
we observed at the level of the hyperfine field are likely to be
related to the transport properties of the material and in
particular to CMR.

Magnetic entities such as magnetic polarons are thought to play a
crucial role in these materials. In particular, they are thought
to be the main cause of CMR. We will show below that a simple
model assuming that magnetic entities are present in the material
and give rise to the static NMR line broadening, which we observe
at higher temperatures, is not consistent with the observed
relaxation times.

Suppose that in order to explain the narrow lines at low
temperatures, we assume the magnetic entities diffuse rapidly at
low temperatures, giving rise to motional narrowing. Thus they
would create a fluctuating field at the nuclear site with
correlation time $\tau$. The relaxation time deriving from the
presence of this fluctuating field could then be written

\begin{equation}
\label{Motional narrowing}
\frac{1}{T_2}=\frac{\delta\omega^2\tau}{1+(\delta\omega\tau)^2}
\end{equation}
where $\delta\omega$ is the distribution in NMR frequency caused
by the fluctuating magnetic field.\cite{Slichter01} Since we
observed that the line is broadened at high temperature, we assume
that at high temperature the correlation time is long, i.e.
$T_2\cong \tau$. This is consistent with the model of magnetic
polarons, since magnetic polarons are thought to be trapped at
high temperature (leading to a decrease in conductivity). At low
temperature we assume that the entities are moving, which means
that $\tau$ is short. In this case, $1/T_2\cong\delta\omega^2\tau$
and the line is motionaly narrowed. In between the high
temperature regime and the low temperature regime, we expect that
$\delta\omega\tau=1$ at a given temperature. At that temperature,
$1/T_2$ is maximum and $T_2=2\tau=2/\delta\omega$. Assuming that
$\delta\omega$ is temperature independent, that means that the
relaxation rate is about equal to the static linewidth at high
temperatures. But in fact, at all temperatures $1/T_2$ is much
less than the observed line width. Thus we cannot explain the data
consistently with this model. Instead, we can argue from NMR that
the material at low temperatures is spatially uniform, but as the
temperature rises, non-uniformities set in.

We also measured the effect of an external magnetic field of 4T on
the temperature dependence of the FWHM of the line. A field of 4T
was strong enough to saturate the magnetization, and as a
consequence a non-negligible demagnetizing field was present in
the non-perfectly ellipsoidal sample. Because of the inhomogeneity
in the demagnetizing field, we measured a FWHM of about 0.5\,MHz
at 10\,K, which was about 2.5 times broader than the zero-field
FWHM at the same temperature. However, the FWHM did not increase
with increasing temperature, and it was still about 0.5\,MHz at
50\,K. We concluded that the temperature dependent static
inhomogeneities that we measured at zero-field were not present
when there was a strong external magnetic field. This observation
is an additional suggestion that the broadening might be linked to
transport properties of the material. Indeed, since the
conductivity of CMR materials is increased by the application of
an external magnetic field, applying a field might perhaps be
viewed as equivalent to lowering the temperature, as far as the
transport properties are concerned.\cite{Shapira01} Therefore, at
a given temperature we can expect the line to be narrower in the
presence of an external field than in zero-field.

\begin{figure}
\includegraphics [width=3.41in] {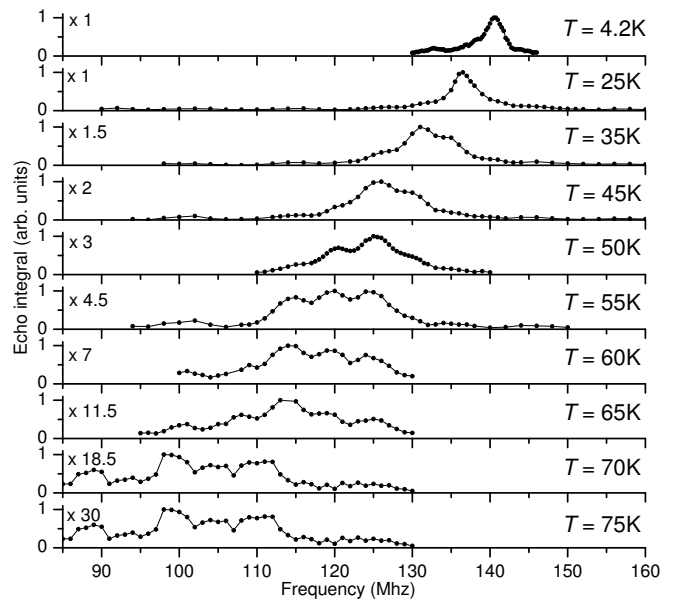}
\caption{\label{fig:LineshapeGd2EuOvsT} Zero-field lineshape of
Eu$^{153}$ in 2\% Gd doped EuO vs. temperature. The amplitude of
the lineshape was multiplied by the coefficient shown on the left
side of each curve.}
\end{figure}

The center of the 2\% and 4.3\% Gd doped EuO lines are
considerably shifted towards higher frequency (140.6\,MHz and
140.0\,MHz respectively). However, we did not observe a monotonic
increase of the shift with the doping, the shift being smaller for
4.3\% Gd doped EuO than for 2\% Gd doped EuO. We also observed
that increasing the Gd doping strongly influences the shape of the
$\nuc{Eu}{153}$ line (c.f. Fig.\,\ref{fig:LineshapeGd2EuOvsT}). In
particular, the lines of 2\% and 4.3\% Gd doped samples were
considerably broader than in the case of pure EuO and 0.6\% Gd
doped EuO and we could not distinguish the central transition from
the other transitions. In order to decide whether or not this
broadening mechanism was magnetic in origin, we measured the
$\nuc{Eu}{151}$ lineshape of the 4.3\% sample and we compare the
$\nuc{Eu}{153}$ and $\nuc{Eu}{151}$ lineshapes. The frequency of
the $\nuc{Eu}{151}$ lineshape was multiplied by the ratio
$^{151}\gamma_n/^{153}\gamma_n$. Since the two curves had a
similar shape, we concluded that the broadening mechanism was of
magnetic origin. We do not know exactly what interaction causes
this large broadening, but we do know from the $T_2$ measurements
(c.f. Sect.\,\ref{sec:T2}) that the broadening is a static
broadening. In conclusion, we observed a remarkable difference
between the lineshape of samples with none or low Gd doping and
the lineshape of samples with higher Gd doping.

We also measured the temperature dependence of the $\nuc{Eu}{153}$
lineshape in 2\% and 4.3\% Gd doped EuO. We plotted the
measurements for 2\% Gd doped EuO in
Fig.\,\ref{fig:LineshapeGd2EuOvsT}. A similar temperature behavior
was observed in 4.3\% Gd doped EuO. We performed point-by-point
measurements to determine the lineshapes. Note that at
temperatures higher than about 80\,K we could not detect the
resonance. This failure was probably due to the fact that the line
was too broad and therefore the signal was too weak.

As shown in Fig.\,\ref{fig:LineshapeGd2EuOvsT}, the lineshape
becomes broader with increasing temperature. Also, the structure
of the lineshape becomes more and more complex with increasing
temperature. In Fig.\,\ref{fig:FWHMvsT} we plotted the FWHM of the
lineshape as a function of temperature for 2\% and 4.3\% Gd doped
EuO along with the temperature dependence of the FWHM of the
lineshape of pure EuO and 0.6\% Gd doped EuO. At temperatures
above about 50\,K the FWHM of 2\% and 4.3\% Gd doped EuO is
ill-defined since the line has a complex shape. We determined that
the temperature dependent broadening was magnetic in origin and
since $1/T_2$ is by far smaller than the linewidth of the lines we
measured, we concluded that the temperature dependent broadening
is a static magnetic broadening.

A possible explanation for the temperature dependent broadening in
2\% and 4.3\% Gd doped EuO is that the value of the exchange
integral is distributed throughout the samples, i.e., different
part of the samples have different values of $J$. For any
reasonable model of M(T), it would take a large change in $T_C$,
hence of J, to account for the spread in frequency we observed.
However, a slight variation of J strongly influences the value of
$T_1$ because $T_1\propto J^{9/2}$. We observed that the measured
$T_1$ was, to a good approximation, independent of the position on
the line. Therefore, we must conclude that there is a negligible
distribution in $J$.

\section{Conclusions}

The study of pure and Gd doped EuO was motivated by the
unconventional electrical and magnetic properties observed in this
system. Our NMR measurements led to results giving new information
on the microscopic magnetic properties of Gd doped EuO. We
observed a dramatic difference between pure or nearly pure (0.6\%
Gd doped) samples, and samples with higher Gd concentration both
in static and dynamic magnetic properties. Below are the main
outcomes of this study.

\subsubsection{Exchange integral J vs. doping}

We observed that the relaxation times in pure and Gd doped EuO
were in good agreement with a law derived from spin-wave theory
for temperatures as high as $T\cong0.6\,T_{\text C}$. From the
temperature dependence of the relaxation times, we inferred the
value of the exchange integral $J$ as a function of Gd doping. We
observed an abrupt increase of the amplitude of the exchange
interaction with doping. This is consistent with the dependence on
Gd concentration of the Curie temperature for which different
theoretical explanations have been proposed by Mauger {\it et
al.},\cite{Mauger02,Mauger03} Nolting and
Ole\'{s},\cite{Nolting01} and more recently Santos and
Nolting.\cite{Santos01}

\subsubsection{Variation of static magnetic broadening with temperature in Gd doped EuO}

We discovered that the lineshape of the Eu resonance in Gd doped
EuO was broader than in pure EuO and that the broadening increases
abruptly with Gd concentration. We also observed that the
broadening increases with increasing temperature. We showed that
the broadening was due to inhomogeneities in the local magnetic
field acting on the Eu nuclei. We confirmed that the increase in
broadening was not due to temperature dependent fluctuations of
the electron spins. $1/T_2$ was shown to be determined by these
fluctuations, but was considerably smaller than the width of the
broadening. Thus, the magnetic inhomogeneities were static at all
temperatures, at least at the time scale of NMR, which is of the
order of several microseconds. Cooper {\it et al.}, who studied Gd
doped EuO by Raman spectroscopy, also observed magnetic
inhomogeneities in Gd doped EuO.\cite{Snow01,Rho01,Rho02} Our
results confirm theirs but we can go further by affirming that the
inhomogeneities are static.

In 0.6\% Gd doped EuO, we observed that above about 30\,K the
magnetic inhomogeneity increases rapidly with increasing
temperature. According to previous transport measurements, the
resistivity of samples containing a similar Gd concentration
increases dramatically above about
30\,K.\cite{Godart01,Samokhvalov02} This suggests that the
broadening mechanism is linked to the change in transport
properties of the sample. Numerous models explain this change in
transport properties by the formation of bound magnetic polarons.
We could rule out the picture according to which the localized
magnetic entities were highly mobile at low temperatures, giving
motionally narrowed lines and high conductivity.

In samples with higher Gd concentration (2\% and 4.3\%), the
observed static broadening is much larger than in 0.6\% Gd doped
EuO. Also, above about 30\,K, some structure develops in the
lineshape along with the increase of magnetic inhomogeneity in the
samples. We showed that this phenomenon was not due to a
distribution in the value of the exchange integral $J$.

We must conclude that at low temperatures the magnetic phase of
the Gd doped samples is fairly homogeneous, and that static
magnetic inhomogeneities are formed when the temperature is
increased. From our results, we deduce that a theory to be valid
has to include a spatially homogeneous and temperature independent
$J$, but include the presence of magnetic inhomogeneities.
Therefore, the explanation of the magnetization curve given by
Borukhovich {\it et al.} is incompatible with our results since it
assumes a distribution of $J$ in the sample.\cite{Borukhovich01}
Although both the theories of Mauger \et.\cite{Mauger03} and
Nolting \et.\cite{Nolting01} assume a spatially homogeneous $J$
and seem to correctly describe the doping dependence of the Curie
temperature, they both suppose that $J$ is temperature dependent,
which in the light of our results cannot be the case. In addition,
these models assume that the magnetization is uniform within the
sample, i.e. they do not take into account any magnetic
inhomogeneities.

There remains the possibility that the charge carrier trapping
above about 30\,K that accounts for the dramatic temperature
dependence of the electrical resistivity causes a distribution in
hyperfine field, presumably because the charge carriers are not
all in the same spin state. This description is compatible with a
theory based on the formation of magnetic polarons taken to be
static on the NMR time scale. However, to the best of our
knowledge, a complete model containing all the key features of the
Gd doped EuO systems does not yet exists.

\begin{acknowledgments}

We are very grateful to Dylan F. Smith for his invaluable help
during the various experiments. We also would like to thank K.
Mattenberger for providing the samples. This work was supported by
the U.S. Department of Energy, Division of Materials Sciences,
through the Frederick Seitz Materials Research Laboratory at the
University of Illinois at Urbana-Champaign under Award No.
DEFG02-91ER45439.
\end{acknowledgments}

\appendix

\section{\label{sec:appendixA}Three-magnon relaxation process}

Below we present the derivation of the formula giving the
relaxation rate due to three-magnon relaxation process for the
case of zero-field NMR in a ferromagnet with low anisotropy. Using
a similar derivation than the one leading to (14) in the paper of
Barak \et., \cite{Barak01} we obtained:

\begin{widetext}
\begin{eqnarray}
\label{Relax rate Barak 4} \lefteqn{\frac{1}{T_1}=\frac{\pi
A^2a^9}{8\hbar
S}\int\limits_0^\infty\int\limits_0^\infty\frac{\exp((\epsilon_{\vec{k}_1}+g\mu_B|\vec{H}|)/k_BT)}{\exp((\epsilon_{\vec{k}_1}+g\mu_B|\vec{H}|)/k_BT)-1}}\nonumber \\
& &
\times\frac{\exp((\epsilon_{\vec{k}_2}+g\mu_B|\vec{H}|)/k_BT)}{\exp((\epsilon_{\vec{k}_2}/k_BT+g\mu_B|\vec{H}|))-1}\;
\frac{g(\epsilon_{\vec{k}_1})g(\epsilon_{\vec{k}_1})g(\epsilon_{\vec{k}_2}+\epsilon_{\vec{k}_2})d\epsilon_{\vec{k}_1}d\epsilon_{\vec{k}_2}}
{\exp((\epsilon_{\vec{k}_1}+\epsilon_{\vec{k}_2}+2g\mu_B|\vec{H}|)/k_BT)-1}.
\end{eqnarray}
\end{widetext}
where $a$ is the lattice constant of the crystal (we used the fact
that EuO and Gd doped EuO are cubic crystals), $g$ is the
g-factor, $\mu_B$ is the Bohr magneton,
$\vec{H}=\vec{H}_{an}+\vec{H}_0+\vec{H}_{dm}$, where
$\vec{H}_{an}$ is the anisotropy field, $\vec{H}_0$ is the applied
external field, and $\vec{H}_{dm}$ is the demagnetization field,
$\epsilon_{\vec{k}_i}$ is the spin-wave energy, and
$g(\epsilon_{\vec{k}_i})$ is the density of states of the spin
waves. We then used the fact that the low energy spin waves give
the principal part of the integral in (\ref{Relax rate Barak 4}),
\cite{Barak01} and we therefore determined
$g(\epsilon_{\vec{k}_i})$ by using the approximation
$\epsilon_{\vec{k}_i}=2JS\vec{k}_i^2a^2$. We obtained:

\begin{widetext}
\begin{eqnarray}
\label{Pincus Beeman before integration}
\lefteqn{\frac{1}{T_{1}}=\frac{1}{16(2\pi)^5}\frac{A^2}{2JS\cdot
\hbar S}\frac{1}{(2JS)^{7/2}}\int\limits_0^\infty\int\limits_0^\infty\frac{\exp((\epsilon_{\vec{k}_1}+g\mu_B|\vec{H}|)/k_BT)}{\exp((\epsilon_{\vec{k}_1}+g\mu_B|\vec{H}|)/k_BT)-1}}\nonumber \\
& &
\times\frac{\exp((\epsilon_{\vec{k}_2}+g\mu_B|\vec{H}|)/k_BT)}{\exp((\epsilon_{\vec{k}_2}/k_BT+g\mu_B|\vec{H}|))-1}\;
\frac{\sqrt{\epsilon_{\vec{k}_1}^2\epsilon_{\vec{k}_2}+\epsilon_{\vec{k}_2}^2\epsilon_{\vec{k}_1}}}
{\exp((\epsilon_{\vec{k}_1}+\epsilon_{\vec{k}_2}+2g\mu_B|\vec{H}|)/k_BT)-1}\;d\epsilon_{\vec{k}_1}d\epsilon_{\vec{k}_2}.
\end{eqnarray}
\end{widetext}

Beeman and Pincus used the fact that $k_BT\gg g\mu_B|\vec{H}|$,
and they found the integral in (\ref{Pincus Beeman before
integration}) to be equal to $7.6\,(k_BT)^{7/2}$ (c.f. (2.22) in
the article of Beeman and Pincus \footnote{There is an error in
equation (2.22) where $h^{7/2}$ should be replaced by
$\hbar^{7/2}$.}).\cite{Beeman01} We recomputed the integral in
order to obtain a value adapted to the case of zero-field NMR
measurements of Eu nuclei in EuO and Gd doped EuO. In that case,
$|\vec{H}|=|\vec{H}_{an}|$ and at 4.2\,K the ratio
$g\mu_B|\vec{H}_{an}|/k_BT$ is approximately equal to
$8\cdot10^{-3}$. Since we observed that the three-magnon process
is the main relaxation process for temperatures between about
15\,K and 50\,K, we had to take a substantially smaller value of
$g\mu_B|\vec{H}|/k_BT$. Taking $T=30$\,K and
$|\vec{H}_{an}(T=30$\,K$)|/|\vec{H}_{an}(T=4.2$\,K$)|\cong2.5$,\cite{Hughes01}
we obtained $g\mu_B|\vec{H}|/k_BT\cong4\cdot10^{-4}$. With this
value, we evaluated the integral to be approximatively
$10.5\,(k_BT)^{7/2}$. This value is very close to the value
obtained if we assume $g\mu_B|\vec{H}|/k_BT=0$, and since there is
no apparent reason to compute the integral with $T=30$\,K instead
of any other temperature in the range 15\,K$<T<$50\,K, we decided
to use the result corresponding to the approximation
$g\mu_B|\vec{H}|/k_BT=0$, that is $11.29\,(k_BT)^{7/2}$. This
result leads to the expression given in (\ref{Oguchi}).

Since we were interested in evaluating $J$ using (\ref{Pincus
Beeman before integration}), we had to determine the influence of
a variation in the value of the integral on $J$. We calculated
that if the factor 11.29 is replaced by the factor 10.5 obtained
for $T=30$\,K, the value of $J$ would be changed by about 1\%.
Therefore we decided that (\ref{Oguchi}) could be used to
determine a fairly accurate value of $J$ from our data.

\bibliography{GdEuO}

\end{document}